%% file: iclr_2023.tex
\documentclass{article} 
\usepackage{iclr2023_conference,times}

\input{math_commands.tex}

\usepackage{hyperref}
\usepackage{url}
\usepackage[utf8]{inputenc} 
\usepackage[T1]{fontenc}    
\usepackage{color}
\definecolor{alice}{RGB}{150, 206, 180}
\definecolor{bob}{RGB}{255, 173, 96}

\usepackage{booktabs} 
\usepackage{url}            
\usepackage{amsfonts}       
\usepackage{nicefrac}       
\usepackage{microtype}      
\usepackage{graphicx}
\usepackage{subfigure}
\usepackage{algorithm}
\usepackage{algorithmic}
\usepackage{braket}
\usepackage{multirow}
\usepackage{multicol}
\usepackage{pythonhighlight}

\usepackage{caption}
\usepackage{mathtools}
\usepackage{lscape}
\usepackage{amssymb}
\usepackage{mathtools}
\usepackage{amsthm}
\usepackage[capitalize,noabbrev]{cleveref}
\usepackage{tikz}

\theoremstyle{plain}

\theoremstyle{definition}

\theoremstyle{remark}

\title{Adapting Pre-trained Language Models for Quantum Natural Language Processing}

\author{Qiuchi Li \\
  University of Copenhagen \\
  {\tt qiuchi.li@di.ku.dk}\\
  \And
  Benyou Wang \\
   the Chinese University of Hong Kong, Shenzhen  \\
  {\tt wangbenyou@cuhk.edu.cn} \\
  \AND
  Yudong Zhu \\
  Baidu  \\
  {\tt zhuyudong3@huawei.com} \\
  \And
  Qun Liu \\
  Huawei Noah's Ark Lab  \\
  {\tt qun.liu@huawei.com} \\
  \And
  Christina Lioma \\
  University of Copenhagen  \\
  {\tt c.lioma@di.ku.dk} \\
  }

\iclrfinalcopy 
\begin{document}

\maketitle

\begin{abstract}
The emerging classical-quantum transfer learning paradigm has brought a decent performance to quantum computational models in many tasks, such as computer vision, by enabling a combination of quantum models and classical pre-trained neural networks. However, using quantum computing with pre-trained models has yet to be explored in natural language processing (NLP). Due to the high linearity constraints of the underlying quantum computing infrastructures, existing Quantum NLP models are limited in performance on real tasks. We fill this gap by pre-training a sentence state with complex-valued BERT-like architecture, and adapting it to the classical-quantum transfer learning scheme for sentence classification. On quantum simulation experiments, the pre-trained representation can bring 50\% to 60\% increases to the capacity of end-to-end quantum models.
\end{abstract}

\section{Introduction}
\label{sec:introduction}

\begin{figure}[ht] 
\centering 
\includegraphics[width=\textwidth]{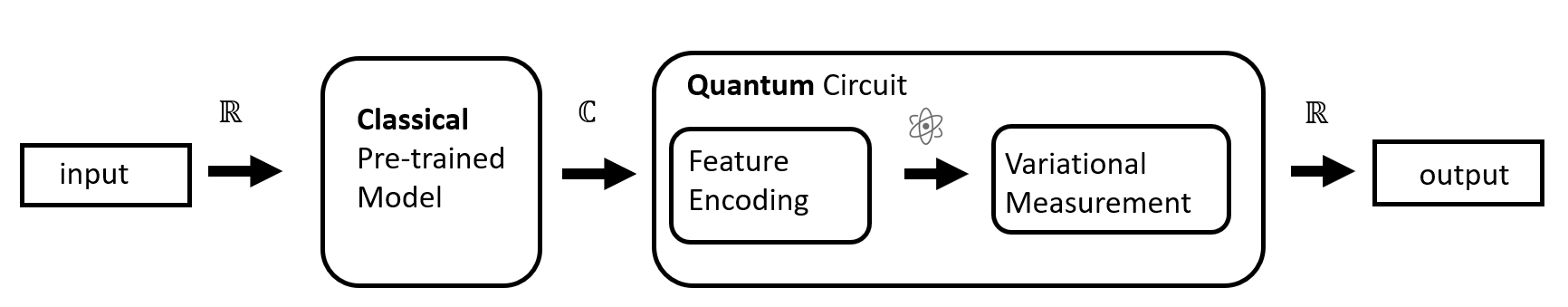}
\caption{Classical-quantum transfer learning pipeline \cite{mari2020transfer}.  }
\label{fig:quantum_transfer}
\end{figure}

Quantum computing combines quantum mechanics and computer science. The concepts of superposition and entanglement bring inherent parallelism between \textit{qubits}, the basic computational element, which endow enormous computational power to quantum devices. 
Classical–quantum transfer learning~\citep{mari2020transfer} has emerged as an appealing quantum machine learning technique. As shown in Fig.~\ref{fig:quantum_transfer}, in a classical–quantum transfer learning pipeline, the pre-trained input features are encoded to a multi-qubit state, transformed and measured in a quantum circuit. The output probabilities are projected to the task label space. The losses are backpropagated to update the parameters in the pipeline with classical algorithms.
This transfer learning scheme combines the representation power of state-of-the-art (SOTA) pre-trained models and the computational power of quantum computing, yielding decent accuracy on various image classification tasks~\citep{lloyd2020quantum,mogalapalli_classicalquantum_2021, mari2020transfer, oh_tutorial_2020}.

However, combining pre-trained models and quantum computing remains unexplored in NLP, where large-scale pre-trained models have dramatically improved language representation~\citep{devlin2019bert,radford2018improving}. Current Quantum NLP (QNLP) models~\citep{zeng2016quantum,coecke2020foundations, meichanetzidis2020quantum, lorenz2021qnlp, lloyd2020quantum, kartsaklis_lambeq_2021} mainly construct quantum circuits from a certain kind of tensor network that aggregates word vectors to sentence representations~\citep{coecke2020foundations}, and the parameters in the network are randomly initialized and learned end-to-end. Since a quantum circuit can be seen as a linear model in the feature space~\citep{schuld_quantum_2021}, these models are highly restricted in scalability and effectiveness. One attempt to resolve this issue is hybrid classical-quantum models~\citep{li_2022_quantum}, where certain layers of models are implemented on a quantum device, and the intermediate results are sent to classical models for non-linear operations. However, the frequent switching between classical and quantum processing units significantly drags the speed of training and inference, and limits the applicability of the model. 

We posit that classical-quantum transfer learning paradigm is a sound fit for QNLP models, especially in the current noisy intermediate-scale quantum (NISQ) era. The pre-trained language features can lead to strong performance in downstream natural language understanding tasks~\cite{devlin2019bert}, even with simple models. Furthermore, it is crucial to introduce robust quantum encodings to mitigate the errors caused by the noisy quantum device, and pre-trained language model is a promising approach to this aim due to its high robustness~\cite{qiu_pre-trained-2021}. Finally, pre-trained mechanism can contribute to scalable QNLP models by avoiding tensor product of all token vectors and using fixed-dimensional representations for arbitrarily long sentences. 

We are motivated to pre-train language representations compatible with quantum computing models. Due to the crucial role of complex numbers in quantum computing, we build a complex-valued pre-trained language model (PLM) for classical-quantum transfer learning. Complex-valued neural networks (NNs) have been long studied~\citep{georgiou1992complex,nitta2002on,hirose2011nature, trabelsi2018deep, xiang2020interpretable, yang2021complex} with various NN building blocks including RNN~\citep{wisdom2016full}, CNN~\citep{guberman2016complex} and Transformers~\citep{yang2021complex, Wang2020encoding, tay_lightweight_2019,zhang_beyond_2021}, and have shown advantages in enhanced representation capacity~\citep{wisdom2016full, arjovsky2016unitary, trabelsi2018deep, Wang2020encoding, trouillon2016complex}, faster learning speed~\citep{arjovsky2016unitary,danihelka2016associative}, and increased model robustness~\citep{danihelka2016associative, yeats21aimproving, xiang2020interpretable}. Despite these advances, complex numbers are not used in pre-trained language models. It remains unknown whether complex-valued NN building blocks can be integrated into high-quality pre-trained models.

To adapt the complex-valued pre-trained models to QNLP, we impose numerical constraints to the network components. For feature encoding, we unit normalize the hidden vectors of the [CLS] token so that the sentence representation can be mapped to a quantum state all throughout the network. We also re-implement the next sentence prediction (NSP) head to simulate variational measurement. At fine-tuning, we train an authentic task-related variational measurement process by parameterizing the involved unitary transformation. Despite the imposed numerical constraints, the quantum-compatible pre-trained language model performs in par to a real-valued BERT of comparable size. More importantly, the pre-trained sentence state encoding brings remarkable performance gain to end-to-end quantum models on various classification datasets, with an relative accuracy improvement of around 50\% to 60\%.

\textbf{We contribute} the first approach to introduce the classical-quantum transfer learning for QNLP models, and pre-train language representations compatible with quantum computing models. Apart from the remarkably improved performance, our model is more scalability then existing NLP models and can tackle longer sentences.




\section{Related work}

\subsection{Complex-valued Neural Networks}
Complex values have been used in various NNs across domains~\citep{arjovsky2016unitary,danihelka2016associative,wisdom2016full,trouillon2016complex,hirose2011nature, trabelsi2018deep, xiang2020interpretable, yang2021complex,guberman2016complex, wang2019semantic}. ~\citet{arjovsky2016unitary,wisdom2016full,wolter2018gated} studied complex numbers in recurrent NNs. ~\citet{arjovsky2016unitary} systematically studied variants of CNNs with complex-valued inputs and weights, which led~\citet{trabelsi2018deep} to build a complex-valued NN that achieved SOTA performance in audio-related tasks. ~\citet{wang2019semantic, yang2021complex, zhang_beyond_2021, tay_lightweight_2019} obtained promising results on sequence-to-sequence (seq2seq) tasks with a complex-valued Transformer. Complex-valued NNs have also been used in privacy detection~\citep{xiang2020interpretable} and knowledge graph completion~\citep{trouillon2016complex, trouillon2017complex}.

Apart from effectiveness gains~\citep{wisdom2016full, arjovsky2016unitary, trabelsi2018deep, Wang2020encoding, trouillon2016complex}, complex-valued NNs also contribute to faster learning~\citep{arjovsky2016unitary,danihelka2016associative} and increased model robustness~\citep{danihelka2016associative, yeats21aimproving, xiang2020interpretable}. However, these properties have been found only in end-to-end tasks. The impact of complex values to pre-trained models remains unexplored.



\subsection{Quantum Natural Language Processing (QNLP)}


QNLP aims to build NLP
models compatible with quantum hardware. Current QNLP models are limited in their architecture and application~\citep{coecke2020foundations, meichanetzidis2020quantum, lorenz2021qnlp, lloyd2020quantum}; they are based on a compositional model~\citep{coecke2010mathematical}, which represents words as tensors in spaces dependent on their grammatical roles, and performs tensor operations to encode syntactic relations. Sentence representations are built by bottom-up aggregating individual word tensors. This process is translated to quantum circuits, followed by a quantum measurement component to produce classification labels. Preliminary studies have successfully implemented and simulated QNLP models for sentence classification~\citep{meichanetzidis2020quantum, lorenz2021qnlp}. By quantum simulation in the feedforward pass and performing backpropagation on a classical computer, the model can learn from data and outperform random labeling.

Like other quantum machine learning models~\citep{lloyd2020quantum, jerbi2021quantum}, QNLP models encode words to different \textit{qubits}, and design the quantum circuit routine (a.k.a \textit{ansatz}) to derive the sentence representation before feeding it to a measurement layer. In its mathematical form, the model encodes each word to a unit complex vector and it has an all-linear structure up to the measurement outcome. Therefore, they have a low capacity and suffer from the scalability issue. Recently, a hybrid classical-quantum scheme~\citep{li_2022_quantum} is introduced to alleviate the limitations. In this quantum self-attention neural network (QSANN), a quantum process is introduced, with parameterized unitary rotations and Pauli measurement, to compute the query, key and value vectors, and they are sent to classical computer to perform non-linear attentions. Due to the introduced non-linearity, QSANN beats the QNLP model in~\citep{lloyd2020quantum}. However, running the network requires switching between quantum and classical hardware at each self-attention layer, which is too inefficient to be practical. We therefore posit that the classical-quantum transfer learning paradigm~\citep{mari2020transfer} is a more promising solution for alleviating the low non-linearity issue for QNLP models.

\section{Background}\label{sec:background_brief}

\noindent\textbf{Complex number.} A complex number $z$ is written as $z =a+bi$ in the rectangular form or $z= re^{i\theta} = r(\cos \theta + i\sin \theta)$ in the polar form. a, b are the real and imaginary parts of $z$, written as $\mathfrak{Re}$(z) and $\mathfrak{Im}$(z). $r=|z| \in [0, +\infty)$ and $\theta \in [-\pi, \pi)$ are the \textit{modulus} (or \textit{amplitude}) and \textit{argument} (or \textit{phase}). The conjugate of a complex number $z = a+bi$ is $\overline{z} = a-bi$, which could be extended for vectors and matrices. The Hermitian of a complex matrix $A$ is its conjugate transpose, written as $A^H = \overline{A^T}$. $A$ is an Hermitian matrix when $A=A^H$. An orthogonal or unitary complex matrix $U$ satisfies $U^HU=I$. We use the typical complex-valued fully-connected (dense) layer to illustrate complex additions and multiplications. A complex dense layer is parameterized by a complex matrix $\mW =\mA + i\mB$ and a complex bias $\vb = \vc + i\vd$. For a complex input vector $\mX =\vx + i\vy$, the output is a complex-valued multiplication of the weight and the input vector, plus the bias value:
\begin{equation} \small
\label{eq:complex_dense}
    {\vz} = \mW \mX + \vb =\left( \mA \vx - \mB \vy + \vc) + i(\mB \vx + \mA \vy + \vd \right) \\
\end{equation}

The \textbf{mean and variance} of a set of complex numbers $\{z_j\}_{j=1}^n$ are given below:

\begin{equation} \small
\label{eq:stats}
    \begin{aligned}
        \bar{\textbf{z}} &= \frac{\sum_{j=1}^n {z_i}}{n} \\
        \sigma^2_z &= \frac{\sum_{j=1}^n {(z_j-\bar{\textbf{z}})(\overline{z_j-\bar{\textbf{z}}})}}{n}.
    \end{aligned}
\end{equation}

\noindent\textbf{Quantum Computing.} We present the basic concepts of quantum computing, see ~\citep{Nielsen2011Quantum} for more. The basic computing unit is a \textit{qubit}, the quantum analog of a classical bit. A qubit describes the state $\ket{\psi}$\footnote{In Dirac's Notations, $\ket{\psi}$ and $\bra{\psi}$ refer to a complex-valued unit row and column vector, respectively.} in a 2-dim Hilbert space. The \textit{basis states} $\ket{0}$ and $\ket{1}$ are orthonormal vectors that form the basis of the space. A general state $\ket{\psi}$ is a \textit{superposition} of the basis states, i.e. $\ket{\psi} = \alpha\ket{0} + \beta \ket{1}$, where $\alpha$, $\beta$ are complex numbers with $|\alpha|^2+|\beta|^2 = 1$. One can apply \textit{measurement} to a qubit to check its probabilities of outcomes 0 and 1 by Born's rule~\citep{born_zur_1926}, i.e, $P(i) = |\braket{\psi|i}|^2$, so $P(0) = |\alpha|^2$ and $P(1) = |\beta|^2$ for the state above. The probabilities of all outcomes always sum to 1. For multiple qubits, their joint space is the tensor product of each qubit space, hence of dimension $2^n$ for $n$ qubits, and the basis states are denoted by $\{\ket{a_1a_2...a_n}, a_i\in \{0,1\}\}$. A state transformation is mathematically a unitary map or a complex unitary matrix $U$, such that $\ket{\psi'} = U\ket{\psi}$ for state $\ket{\psi}$. Quantum circuits are physical implementations of quantum computing models. The basic units of quantum circuits are quantum gates, which are unitary maps that play similar roles to logic gates in classical computers. The combination of different quantum states allows us to implement any unitary transformation before the final measurement step.

\noindent\textbf{Classical BERT.} Bidirectional Encoder Representations from Transformers (BERT)~\citep{devlin2019bert}  takes as input a concatenation of two segments (sequences of tokens). The inputs are passed through an embedding layer that adds positional embedding, token embedding and segment embedding. The embeddings are then fed into a stack of $N$ transformer layers, and each layer has a multi-head attention module to enact token-level interactions. The last hidden units of each sequence are used to perform Mask Language Model (MLM) and Next Sentence Prediction (NSP).  The MLM objective $\mathcal{L}_{MLM}$ is a cross-entropy loss on predicting the randomly masked tokens in the sequence, while the NSP loss $\mathcal{L}_{NSP}$ produces binary cross-entropy loss on predicting whether the two segments follow each other in the original text. The overall objective of BERT is $\mathcal{L}_{BERT} = \mathcal{L}_{MLM}+ \mathcal{L}_{NSP}$. BERT is pre-trained on large text corpora, e.g., BOOKCORPUS and the English WIKIPEDIA~\citep{devlin2019bert}. The model is fine-tuned on different text classification and natural language inference datasets, such as the famous GLUE benchmark~\citep{wang2019glue}. The effectiveness on these datasets indicates the performance of the pre-trained model. 

\input{src/cvbert}

\input{src/qbert}

\section{Conclusion}
We have presented the first classical-quantum transfer learning scheme for quantum natural language processing (QNLP). By delicately designing the pre-trained model architecture, we managed to leverage the strong pre-trained language models for enhancing the capacity of QNLP. Empirical evaluation suggests that 50\% to 60\% improvement in effectiveness can be brought about by the pre-trained quantum language encoding. Besides the main finding, we proposed the first AdamW optimizer for training complex-valued BERT models, and we believe it will be beneficial for training other complex-valued neural networks.

The applicability of our model is limited in that current quantum technology cannot support authentic classical-quantum hybrid training for fine-tuning QBERT on downstream NLP tasks. However, we believe that the classical-quantum transfer mechanism and pre-trained models is necessary for scalable QNLP models, and this work has made the crucial first step by demonstrating the enormous potential of pre-trained quantum encodings. We expect that future advances on quantum technologies will make our approach feasible on real quantum computers.

\bibliography{mybibs.bib}
\bibliographystyle{iclr2023_conference}

\newpage
\appendix
\onecolumn
\section{Demonstration of implementing the QBERT fine-tuning structure with a quantum circuit}~\label{sec:app_qiskit}
\noindent We examine whether our QBERT fine-tuning structure can be implemented with a quantum circuit. First, we build the fine-tuning network in PyTorch, which contains a unitary layer, a measurement layer and a linear projection layer. We then feed the model with random pure states and take the output logits. Next, we design a quantum circuit with the network weights using the qiskit toolkit. The same input states for the classical model are taken to initialize the quantum circuit state. With the circuit design and initial state, we are able to compile the circuit and run quantum simulation. The statistics of the output state is taken over a large number of simulations. The obtained occurrences are converted to probabilities of each basis state, and the probabilities are multiplied by the classical linear transformation matrix to produce the predicted logits. We compute the differences between the logits under both networks to judge if they produce identical results.

We run a simulation with $n=3$ qubits, and $k=2$ classes. That means our QBERT fine-tuning structure involves an 8-dim unitary transformation layer, a measurement along the 8 basis states $\{\ket{000},...,\ket{111}\}$, and a linear projection with a 8-by-2 matrix to produce the logits. We randomly initialize a total number of M=16 pure states. The quantum circuit is simulated for a total number of N=100000 times. We compute the mean squared error (MSE) between the outputs under classical and quantum implementations. As a result, we obtain an MSE value of $e=4.29\times 10^-10$. This proves that the QBERT fine-tuning head can be precisely converted to a quantum circuit. In another word, the whole fine-tuning procedure can potentially be instrumented by a hybrid classical-quantum network, which will be explored in future works.

Below are the main body of the codes for implementing the classical and quantum networks as well as running the simulation on random examples.
\begin{python}
# Quantum Implementation
class ParamCircuit(object):
    """ 
    This class provides a simple interface for interaction 
    with the quantum circuit 
    """
    def __init__(self, n_qubits,
                 backend, 
                 unitary_matrix, 
                 tranformation_matrix,
                 shots=1000):
     
        self.n_qubits = n_qubits
        self.all_qubits = [i for i in range(n_qubits)]
        self.unitary_gate = UnitaryGate(unitary_matrix)
        self.backend = backend
        self.shots = shots
    
    def run(self, initial_states):
        all_probs = []
        for state in initial_states:
            self._circuit = qiskit.QuantumCircuit(self.n_qubits)
           self._circuit.initialize(state, self._circuit.qubits)
           self._circuit.append(self.unitary_gate,self.all_qubits)
            self._circuit.measure_all()
            t_qc = transpile(self._circuit,
                            self.backend)
            qobj = assemble(t_qc,
                   shots=self.shots)
            job = self.backend.run(qobj)
            result = job.result().get_counts()
            counts = [result[k] for k in sorted(result)]
            # Compute probabilities for each state
            probs = np.array(counts) / self.shots
            all_probs.append(probs)
        # Get state expectation
        
        return all_probs

# Classical Implementation
from torch.nn import CrossEntropyLoss, MSELoss

import torch.nn as nn
from layers.bert_dense_factory import Unitary
class QBertCLSHead(nn.Module):
    def __init__(self, dim,num_labels=2):
        super(QBertCLSHead, self).__init__()
        self.num_labels = num_labels
        self.dim=dim
        self.pooler = Unitary(self.dim, 
                  init_mode='normal',
                       real_weight=False)

        self.classifier = nn.Linear(self.dim, self.num_labels, bias=False)
        self.pooler.init_params(std=0.001)
        nn.init.normal_(self.classifier.weight, mean=0, std=0.001)
    
    def forward(self, input_state, labels=None):
    
        pooled_output = self.pooler(input_state)
        
        logits = self.classifier(pooled_output.abs().pow(2))
        
        if labels is not None:
            if self.num_labels == 1:
                #  We are doing regression
                loss_fct = MSELoss()
                loss = loss_fct(logits.view(-1), labels.view(-1))
            else:
                loss_fct = CrossEntropyLoss()
                # print(logits)
                loss = loss_fct(logits.view(-1, self.num_labels), labels.view(-1))
            return loss, logits
        return logits
        
# Running Simulation on Toy Examples 
from models.qiskit.param_circuit import ParamCircuit
from models.qiskit.qcls_head import QBertCLSHead
import qiskit
import torch
import torch.nn.functional as F
import numpy as np
from torch.nn import MSELoss
def run_simulation(qbits=3, batch_size=16, nclasses = 2,shots=1000000):
    
    #qbits = 5
    model_dim = pow(2,qbits)
    qbert_head = QBertCLSHead(model_dim, nclasses)
    r = torch.rand(batch_size, model_dim,dtype = torch.cdouble)
    initial_states = F.normalize(r,p=2, dim=-1)
    outputs = qbert_head(initial_states.to(torch.cfloat))
    unitary_matrix = qbert_head.pooler.compute_unitary().detach()
    backend = qiskit.Aer.get_backend('aer_simulator')
    circuit = ParamCircuit(qbits,backend,unitary_matrix,shots)
    res = circuit.run(initial_states.numpy())
    res = torch.Tensor(np.stack(res,axis=0))
    output = torch.matmul(res,qbert_head.classifier.weight.t())
    print(MSELoss()(output, outputs))
\end{python}

\section{Codes for training DisCoCat on SST with Lambeq}~\label{sec:discocat_sst2}
\begin{python}
# -*- coding: utf-8 -*-

import torch
import pickle
import os
from discopy import Dim
from datasets import load_dataset
from lambeq import AtomicType, SpiderAnsatz, PytorchTrainer, Dataset, PytorchModel, BobcatParser

print('running discocat classification on sst...')
BATCH_SIZE = 32
EPOCHS = 10
LEARNING_RATE = 1e-3

# Otherwise there will be weird bugs in converting the diagrams to tensor networks
model_dim = 2
nb_classes = 2

SEED = 0

max_train_samples = 10000
max_val_samples = 10000

dataset = load_dataset('glue','sst2')

max_train_samples = min(max_train_samples, len(dataset['train']['sentence']))
max_val_samples = min(max_val_samples, len(dataset['validation']['sentence']))

parser = BobcatParser(verbose='text')

train_diagrams = parser.sentences2diagrams(dataset['train']['sentence'][:max_train_samples])
val_diagrams = parser.sentences2diagrams(dataset['validation']['sentence'][:max_val_samples])

train_labels=  [[float(y), float(1-y)] for y in dataset['train']['label'][:max_train_samples]]
val_labels = [[float(y), float(1-y)] for y in dataset['validation']['label'][:max_val_samples]]

max_train_samples = min(max_train_samples, len(train_diagrams))
max_val_samples = min(max_val_samples, len(val_diagrams))

new_train_diagrams = []
new_train_labels = []
for i, (diagram, label) in enumerate(zip(train_diagrams, train_labels)):
    if i >=max_train_samples:
        break
    # Some sentences are not grammatically correct, 
    # and the parser will return None result
    if diagram is not None:
        new_train_diagrams.append(diagram)
        new_train_labels.append(label)
del train_diagrams  
del train_labels

new_val_diagrams = []
new_val_labels = []    
for i, (diagram, label) in enumerate(zip(val_diagrams, val_labels)):
    if i >=max_val_samples:
        break
    if diagram is not None:
        new_val_diagrams.append(diagram)
        new_val_labels.append(label)
    
del val_diagrams
del val_labels

# Convert the diagrams to circuits
ansatz = SpiderAnsatz({AtomicType.NOUN: Dim(model_dim),
                      AtomicType.SENTENCE: Dim(nb_classes),
                       AtomicType.PREPOSITIONAL_PHRASE: Dim(model_dim)})

train_circuits = []
val_circuits = []
legal_ids = []
for i, diagram in enumerate(new_train_diagrams):
    # Some diagrams cannot be converted to circuits,
    # and None will be returned
    try:
        circuit = ansatz(diagram) 
        train_circuits.append(circuit)
        legal_ids.append(i)
    except:
        continue
train_labels = [new_train_labels[x] for x in legal_ids]
legal_ids = []
for i, diagram in enumerate(new_val_diagrams):
    try:
        circuit = ansatz(diagram) 
        val_circuits.append(circuit)
        legal_ids.append(i)
    except:
        continue
val_labels = [new_val_labels[x] for x in legal_ids]

all_circuits = train_circuits + val_circuits 
model = PytorchModel.from_diagrams(all_circuits)

def accuracy(y_hat, y):
    return sum(torch.argmax(y_hat,dim=-1) == torch.argmax(y, dim=-1))/len(y)  

eval_metrics = {"acc": accuracy}

# Create Trainer and datasets from the pre-processed circuits
trainer = PytorchTrainer(
        model=model,
        loss_function=torch.nn.CrossEntropyLoss(),
        optimizer=torch.optim.AdamW,
        learning_rate=LEARNING_RATE,
        epochs=EPOCHS,
        evaluate_functions=eval_metrics,
        evaluate_on_train=True,
        verbose='text',
        seed=SEED)

print('creating dataset...')
train_dataset = Dataset(
            train_circuits,
            train_labels,
            batch_size=BATCH_SIZE)

val_dataset = Dataset(val_circuits, val_labels, shuffle=False)

# Training the model
print('training...')
trainer.fit(train_dataset, val_dataset, evaluation_step=1, logging_step=1)

import matplotlib.pyplot as plt
fig1, ((ax_tl, ax_tr), (ax_bl, ax_br)) = plt.subplots(2, 2, sharey='row', figsize=(10, 6))

ax_tl.set_title('Training set')
ax_tr.set_title('Development set')
ax_bl.set_xlabel('Epochs')
ax_br.set_xlabel('Epochs')
ax_bl.set_ylabel('Accuracy')
ax_tl.set_ylabel('Loss')

colours = iter(plt.rcParams['axes.prop_cycle'].by_key()['color'])
ax_tl.plot(trainer.train_epoch_costs, color=next(colours))
ax_bl.plot(trainer.train_results['acc'], color=next(colours))
ax_tr.plot(trainer.val_costs, color=next(colours))
ax_br.plot(trainer.val_results['acc'], color=next(colours))

# print validation accuracy
test_output = model(val_circuits)
test_labels = torch.tensor(val_labels)
if type(test_output) == tuple:
    test_output, legal_idx = test_output
    test_labels = test_labels[legal_idx]
test_acc = accuracy(test_output, test_labels)
with open('acc_discocat_sst2.txt','w') as writer:
    writer.write(str(test_acc.item()))
\end{python}

\end{document}

%% file: math_commands.tex
\usepackage{amsmath,amsfonts,bm}

\def\eqref#1{equation~\ref{#1}}

\def\1{\bm{1}}

\def\vb{{\bm{b}}}
\def\vc{{\bm{c}}}
\def\vd{{\bm{d}}}

\def\vx{{\bm{x}}}
\def\vy{{\bm{y}}}
\def\vz{{\bm{z}}}

\def\mA{{\bm{A}}}
\def\mB{{\bm{B}}}

\def\mW{{\bm{W}}}
\def\mX{{\bm{X}}}

\DeclareMathAlphabet{\mathsfit}{\encodingdefault}{\sfdefault}{m}{sl}
\SetMathAlphabet{\mathsfit}{bold}{\encodingdefault}{\sfdefault}{bx}{n}



%% file: src/cvbert.tex
\section{Our quantum-compatible pre-trained language model} 
\label{sec:methodology}

\begin{figure*}[t] 
\centering 
\vspace{-15pt}
\includegraphics[width=\textwidth]{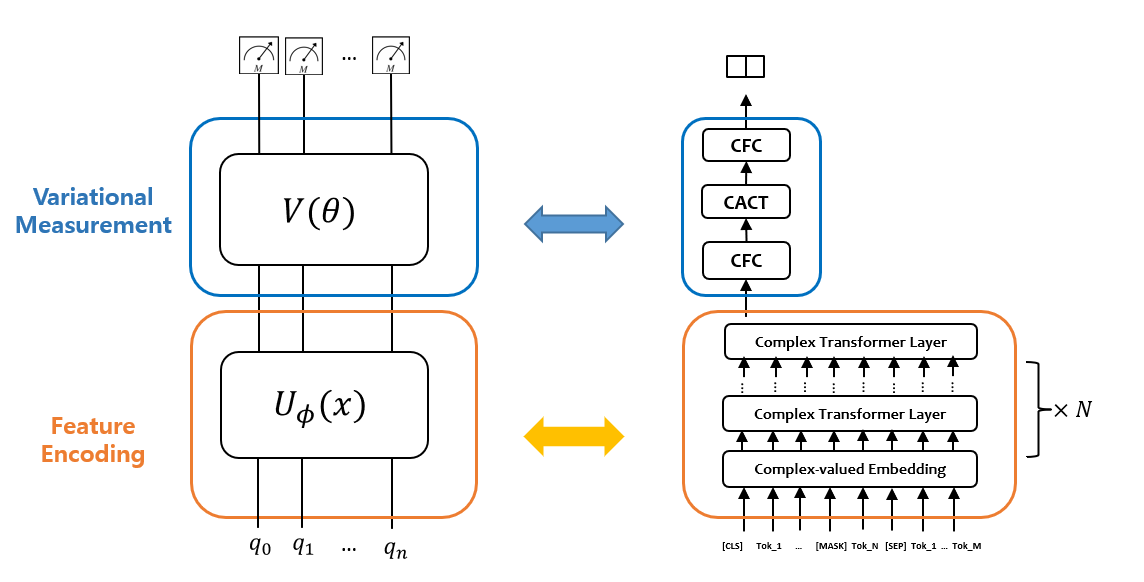}
\caption{Mapping the classical-quantum transfer learning scheme to our complex-valued PLM. The MLM prediction head for PLM and the linear projection layer of the quantum model are omitted.  }
\label{fig:qnn}
\end{figure*}

\subsection{Overall Architecture}

A typical quantum circuit for classification is composed of a feature encoding module and a variational measurement component. The feature encoding unit encodes an input data point $x$, such as a sentence, to any $n$-qubit state $\ket{\psi(x)}$. Basically, it applies a unitary transformation $U_\phi(x)$ to the $n$-qubit basis state, i.e. $\ket{\psi(x)} = U_\phi(x)\ket{00...0}$. Next, the encoded state $\ket{\psi(x)}$ is fed to a variational measurement unit, which consists of a task-related unitary transformation $V(\theta)$ and a measurable observable along the basis states $\{\ket{e_j}\}$ of the $n$-qubit system. The output probabilities on all basis states $p_j = |\braket{e_j|V(\theta)|\psi(x)}|^2$ are aggregated as the circuit output. They are further projected onto a low-dimensional space to produce the task label. 

We establish that the above schema can be structurally mapped to a pre-trained language model, as shown in Fig.~\ref{fig:qnn}. The [CLS] token vector is usually viewed as the sequence representation. The multi-layer Transformer encoder encodes its input to a quantum state $\bm{\ket{\psi_{[CLS]}(x)}}$, playing the role of quantum feature encoding in a quantum circuit. Essentially, the unitary transformation $U_\phi(x)$ is parameterized such that the transformed state $ U_\phi(x)\ket{00...0} =\ket{\psi_{[CLS]}(x)}$. Furthermore, the NSP prediction head that maps [CLS] token to classification label is analogous to the variational measurement component that directs $\bm{\ket{\psi_{[CLS]}(x)}}$ to the classification label. 

Since complex values are vital to quantum models, we build a pre-trained LM with complex values, namely QBERT, to support the mapping above. We follow the multi-layer bidirectional Transformer architecture of classical BERT, and adjust the implementations of each network component to support complex representations. For feature encoding, we unit normalize the hidden vectors of the [CLS] token so that the sentence representation can be attributed to a quantum state at each layer of the network. We also re-implement the NSP prediction head to simulate variational measurement in both pre-training and fine-tuning phase.

\subsection{QBERT building blocks}\label{sec:building_blocks}
\noindent \textbf{Embedding layer.} In classical BERT,  token embeddings, segment embeddings and position embeddings are summed up at a per-token level. They are each extended to the complex domain in QBERT. Essentially, the complex-valued token embeddings, segment embeddings and position embeddings are summed up as the output of the embedding layer for each token.

\noindent \textbf{Multi-head attention.} A complex transformer layer has a multi-head attention component at its core. It computes query-key affinity scores, and linearly combines them with value vectors to produce a contextual vector for each element. The attention scores for one head are computed by

\begin{equation}\small
    \text{ComplexAttention}(Q,K,V) = f(\frac{QK^H}{\sqrt{d_k}})V,
\end{equation}
where $f(\cdot)$ is a softmax-like activation function applied on the query-key complex inner (i.e. Hermitian) products. Since the inputs to this function are complex matrices, we extend the real softmax function to the complex domain.

A straightforward approach to this aim is to apply softmax to real and imaginary parts of the inner product separately. Suppose the Hermitian product is denoted by $\{\sigma(q,k)\}$ for a pair of query-key elements $(q,k)$, the formula of this \textit{split activation function} is given by

\begin{equation}\label{eq:app_split_softmax}
     f_\text{split}(q,k) = \frac{e^{\mathfrak{Re}(\sigma(q,k))}}{\sum_{k'}{e^{\mathfrak{Re}(\sigma(q,k'))}}} +i \frac{e^{\mathfrak{Im}(\sigma(q,k))}}{\sum_{k'}{e^{\mathfrak{Im}(\sigma(q,k'))}}},
\end{equation}

where the summation iterates over all key elements $k'$ in the sequence. The split softmax function normalizes  both real and imaginary parts of the affinity scores to sum up to 1 for each key. When the scores are taken to linearly combine the value vectors $\{v'\}$, the summation can be decomposed into

\begin{equation}
    \resizebox{0.48\textwidth}{!}{
    $
        \begin{aligned}
         h = \sum_{k'} (\mathfrak{Re}(f_\text{split}(q,k'))+i\mathfrak{Im}(f_\text{split}(q,k'))(\mathfrak{Re}(v') + i\mathfrak{Im}(v'))  \nonumber \\
         = \sum_{k'} (\mathfrak{Re}(f_\text{split}(q,k'))\mathfrak{Re}(v')-\mathfrak{Im}(f_\text{split}(q,k'))\mathfrak{Im}(v')) \\ +i\sum_{k'} (\mathfrak{Re}(f_\text{split}(q,k'))\mathfrak{Im}(v')+\mathfrak{Re}(f_\text{split}(q,k'))\mathfrak{Im}(v')),
        \end{aligned} 
    $
    }
\end{equation}

\noindent and a negative sign exists in the real part of the summation due to the $i^2 = -1$. However, an ideal weighted combination should be a convex combination of the value vectors with all non-negative weights. This motivates us to modify the normalization function $f$ so that real-valued affinity scores are produced from the complex inputs, because they indicate a convex combination in both the real and imaginary channels. We propose to apply softmax normalization to the modulus part of the complex numbers, as shown in the equations below.

\begin{equation}\label{eq:app_mod_softmax}
     f_\text{mod}(q,k) =\frac{e^{|\sigma(q,k)|}}{\sum_{k'}{e^{|\sigma(q,k')|}}}
\end{equation}

\noindent \textbf{Feed-forward network.} The main component of feed-forward networks is the fully-connected layer. We employ Eq.~\ref{eq:complex_dense}, Sec.~\ref{sec:background_brief} for The implementation of a complex fully-connected layer. For an input vector $\bm{X} \in \mathbb{C}^{d_i}$, a fully-connected layer projects it to a $d_o$-dim vector $z \in \mathbb{C}^{d_o}$, with weight matrix $\bm{W} \in \mathbb{C}^{d_o \times d_i}$ and a bias term $\bm{b} \in \mathbb{C}^{d_o}$. \\

\textbf{Prediction heads.}  In order to be consistent with the quantum classification model, we re-implement the NSP head to simulate a variational measurement, which consists of unitary training and measurement.  Due to the computational cost of training a unitary matrix, we replace the unitary transformation with a dense layer followed by unit-normalization in the pre-training phase~\citet{chen2021quantum}. Forr the measurement, two pure states are for the NSP task, and the squared Hermitian product between the input state and each state is computed and linearly re-scaled to discrete probabilities. As per~\citet{lorenz2021qnlp}, the probabilities are used to compute the binary cross-entropy loss against the true binary label. We further remove the non-linear activation function from the NSP prediction head.

For MLM head, a complex-valued feed-forward network is used to project the encoder-output tokens to MLM logits, which are then converted to real values by taking the moduli of complex numbers.\\

\textbf{Activation function.}\label{sec:activation} Classical BERT typically adopts Rectified Linear Unit (ReLU)~\citep{relu} or the Gaussian Error Linear Unit (GeLU)~\citep{gelu} as the activation function for hidden units. To extend it to the complex domain, we simply employ \textbf{split-GeLU}, which activates the real and imaginary parts of the input with a GeLU function:

\begin{equation}\label{eq:split_gelu} \small
    \text{split-GeLU}(z) = \text{GeLU}(\mathfrak{Re}(z)) + i \text{GeLU}(\mathfrak{Im}(z))
\end{equation}

\textbf{Layer normalization.}\label{sec:normalization} For a real vector, standard layer normalization rescales the elements to zero mean and unit variance, and applies an affine transformation to the rescaled values. Similarly, one can directly compute the mean $\bar{z}$ and variance $\sigma_z$ for a set of complex numbers $\bm{z}=\{z_j\}_{j=1}^n$ by Eq.~\ref{eq:stats}. The complex layer normalization function becomes 
\begin{equation}\label{eq:complex_layer_norm} \small
    \text{complex-LN}(z) = \frac{z-\bar{z}}{\sigma_z} \times a + b.
\end{equation}

Complex-LN is slightly different from applying layer normalization respectively to real and imaginary channels. They both normalize the mean value of inputs to zero, but bring different variances to the normalized values, and hence lead to different outputs. To ensure the [CLS] token is a legal quantum state, we unit-normalize the hidden vector of the [CLS] token and apply complex-LN to the remaining tokens.

\subsection{Network Training}\label{sec:optim}

\paragraph{Optimization.} Most recent works~\citep{li2019cnm, yang2021complex, tay_lightweight_2019} implement complex-valued NNs with double-sized real networks, and apply classical backpropagation to update their real and imaginary parts simultaneously. However, because of non-holomorphic functions~\citep{hirose2003complex}, this can yield wrong gradients of complex weights, and the effectiveness metrics of a complex-valued NN may not reflect its true performance. Therefore, we use the Wirtinger Calculus~\citep{kreutz-delgado_complex_2009} to update complex-valued parameters, which explicitly computes the gradient with respect to each complex weight. We are the first to adapt AdamW~\citep{loshchilov2018fixing}, the most popular optimizer, for complex weights. AdamW computes the second raw moment (i.e., uncentered variance) of the gradient by averaging the squared gradients in the real case. In the complex domain, however, the variance should be computed by \textit{multiplying the gradient with its conjugate}. We modify AdamW accordingly to fix this problem and get the correct second raw moment for complex gradients. We highlight the difference between the real and complex AdamW optimizers in Alg.\ref{alg:complex_adamw}. 

\noindent \textbf{Weight Initialization}. By default, we initialize the real and imaginary parts of complex weights
with a normal distribution at a mean value of zero and a variance of 0.01.

\subsection{Measurement as Classification}

With the pre-trained quantum encoding $\bm{\ket{\psi_{[CLS]}(x)}}$, we train a task-related variational measurement for text classification. The state is passed to an authentic unitary layer, parameterized by a complex-valued square matrix $W$ as follows:
\begin{equation} \label{eq:unitary} \small
 H = \frac{W + W^H}{2}, \; \; U = e^{iH},
\end{equation}
where $e^{(\cdot)}$ stands for matrix exponential. The resulting $U$ is guaranteed to be a square unitary matrix, supporting a trainable unitary transformation of [CLS] state $\bm{\ket{\psi'_{[CLS]}(x)}} = U\bm{\ket{\psi_{[CLS]}(x)}}$. Finally, $\bm{\ket{\psi'_{[CLS]}(x)}}$ is measured along each basis state, and the output probability vector is linearly transformed to produce the target class labels. The variational measurement head is trained with the previous transformer layers in a fine-tuning task. 

At this stage, the model operates in a hybrid classical-quantum fashion: the multi-layer Transformer encoder is executed in a classical computer to obtain the encoded state $\bm{\ket{\psi_{[CLS]}(x)}}$. The state is then passed to a quantum device to compute the classification probabilities for each class. Finally, we compute the cross-entropy loss against true class labels as the loss function, and the network weights are updated accordingly with the \textbf{CAdamW} optimizer in a classical computer. Compared to QSANN~\citep{li_2022_quantum}, this model only requires switching once between quantum and classical hardware, so it is far more practical for hybrid classical-quantum training.

\begin{algorithm}[t] \small
    \caption{AdamW \textcolor{alice}{for real numbers} and AdamW for \textcolor{bob}{for complex numbers}}
    \label{alg:complex_adamw}
    \begin{algorithmic}[1]
        \STATE \textbf{Given} $\alpha =0.001$, $\beta_1=0.9$, $\beta_2=0.999$, $\epsilon=10^{-8}$, $\lambda \in \mathbb{R}$
        \STATE \textbf{Initialize} time step $t \leftarrow \textbf{0}$, \textcolor{alice}{parameter $\boldsymbol{\theta}_{t=0} \in \mathbb{R}^n$}, \textcolor{bob}{parameter $\boldsymbol{\theta}_{t=0} \in \mathbb{C}^n$}, first moment $\textbf{m}_{t=0} \leftarrow \textbf{0}$, second moment $\textbf{v}_{t=0} \leftarrow \textbf{0}$, schedule multiplier $\eta_{t=0} \in \mathbb{R}$
        \WHILE{stopping criteria is not met}             
            \STATE $t \leftarrow t+1$
            \STATE $\bigtriangledown f_t(\boldsymbol{\theta}_{t-1}) \leftarrow \text{SelectBatch}(\boldsymbol{\theta}_{t-1})$
            \STATE $\textbf{g}_t \leftarrow f_t(\boldsymbol{\theta}_{t-1}) + \lambda \boldsymbol{\theta}_{t-1}$
            \STATE $\textbf{m}_t \leftarrow \beta_1 \textbf{m}_{t-1} + (1-\beta_1) \textbf{g}_t $
            \STATE $\textcolor{alice}{\textbf{v}_t \leftarrow \beta_2 \textbf{v}_{t-1} +(1-\beta_2) \textbf{g}_t \odot \textbf{g}_t} $
            \STATE $ \textcolor{bob}{\textbf{v}_t \leftarrow \beta_2 \textbf{v}_{t-1}+(1-\beta_2) \textbf{g}_t \odot \bar{\textbf{g}_t}}$
            \STATE $\hat{\textbf{m}_t} \leftarrow \textbf{m}_t/(1-\beta_1^t)$
            \STATE $\hat{\textbf{v}_t} \leftarrow \textbf{v}_t/(1-\beta_2^t)$
            \STATE $\eta_t \leftarrow \text{SetScheduleMultiplier}(t)$
            \STATE $\boldsymbol{\theta}_t \leftarrow \boldsymbol{\theta}_{t-1} - \eta_t(\alpha \hat{\textbf{m}_t}/(\sqrt{\hat{\textbf{v}_t}}+\epsilon)+\lambda \boldsymbol{\theta}_{t-1})$
        \ENDWHILE
        \STATE \textbf{Return} $\boldsymbol{\theta}_t$
  \end{algorithmic}
\end{algorithm}

%% file: src/qbert.tex
\section{Experiment}

We pre-train QBERT on BOOKCORPUS and English WIKIPEDIA and evaluated them on the GLUE benchmark, following the standard practice. To examine its effectiveness, we compare it with the classical BERT (a.k.a \textbf{BERT-base}) and a complex-valued BERT (a.k.a \textbf{CVBERT-base}). \textbf{CVBERT-base } has the same settings as \textit{QBERT-base}, except that the [CLS] token is layer normalized by \textit{complex-LN}, and the NSP head is not replaced by quantum-compatible structures in pre-training and fine-tuning. We plot their learning curves during pre-training, and compute their effectiveness on GLUE datasets. Following the established practice, \textbf{GLUE scores} are calculated by averaging the performance values on all GLUE datasets~\cite{wang2019glue}.

To ensure a fair comparison, we align the parameter number of all three models. The models have a 12-layer Transformer structure with 12 heads in each layer. \textbf{BERT-base} has a model dimension $d_{model}=768$ and a hidden size of $d_{hidden}=3072$. Since a complex-valued NN has twice the number of parameters as its real-valued counterpart, the complex-valued models \textbf{CVBERT-base} and  \textbf{QBERT-base} have halved hidden dimension $d_{hidden}=1536$ and same model dimension $d_{model}=768$. In this way, a 768-dim unit complex vector is pre-trained  as the quantum encoding $\bm{\ket{\psi_{[CLS]}(x)}}$. This means that the quantum classification model can be implemented as a 10-qubit circuit\footnote{Since the dimensionality is not a power of 2, the capacity of qubits is not fully exploited. We aim at conducting a fair comparison with the other BERT models.}. We further remove the query projection and output projection layers $W^Q$, $W^O$ from all its transformer layers, and tie the input embedding lookup table with the MLM projection matrix. As shown in Tab.~\ref{tab:qBERT}, the real, complex and quantum-compatible BERT models are comparable in size.  

To empirically check the gain in representation capacity brought about by the classical-quantum transfer learning paradigm, we compare the fine-tune performance of \textbf{QBERT-base} with \textbf{DisCoCat}~\cite{lorenz2021qnlp}, the architecture for QNLP models. However, due to the scalability issue, \textbf{DisCoCat} can only work at a low dimensionality to be able to handle long sentences in the datasets. We follow the original setting in~\cite{lorenz2021qnlp}, which uses at most 3 qubits to represent each token. We also implement two other end-to-end \textit{quantum-like}~\footnote{They are not strict quantum models, since the sentence encoding in \textbf{QBERT-base} and attention components in the \textbf{QCLS-transformers} must be implemented on a classical computer. However, they are good estimates of how a quantum model of comparable size to \textbf{QBERT-base} performs on text classification.} models to simulate the performance of quantum models at a comparable scale to \textbf{QBERT-base}. \textbf{QCLS-end2end} model embeds each word to a complex-valued vector and normalizes the average of word vectors as the sequence encoder. The sequence state is passed to a variational measurement to produce classification labels. The \textbf{QCLS-transformer} model has a similar structure to \textbf{QBERT-base} but is directly trained on text classification dataset with randomly initialized complex-valued token embeddings. We try $N\in \{3,6,12\}$ transformer layers in the \textbf{QCLS-transformer} architecture. In the absence of the source code of \textbf{QSANN}~\cite{li_2022_quantum}, the performance of \textbf{QCLS-transformer} can serve as an estimate of \textbf{QSANN}, since they have similar attention mechanisms.

\begin{table*}[ht!]

\caption{Results of \textbf{QBERT-base} on GLUE in comparison to classical BERT models and end-to-end QNLP models. The evaluation metric for each dataset is shown in the parentheses below the dataset name. Performance values on the development set are reported. We report the relative differences between \textbf{QBERT-base} and each model in the parentheses in the last column.}
\label{tab:qBERT}
\resizebox{\textwidth}{!}{
\begin{centering}
\begin{tabular}{l|c|c|c|c|c|c|c|c|c|c|c|c}
\hline\hline

\multicolumn{1}{c|}{Name} & \multicolumn{1}{c|}{$d_{model}$} &\multicolumn{1}{c|}{$d_{hidden}$} &\multicolumn{1}{c|}{Size} & \begin{tabular}[c]{@{}c@{}}MNLI\\(Acc)\end{tabular} & \begin{tabular}[c]{@{}c@{}}QNLI\\ (Acc)\end{tabular} & \begin{tabular}[c]{@{}c@{}}QQP\\ (F1)\end{tabular} & \begin{tabular}[c]{@{}c@{}}RTE\\ (Acc)\end{tabular} & \begin{tabular}[c]{@{}c@{}}SST\\ (Acc)\end{tabular} & \begin{tabular}[c]{@{}c@{}}MRPC\\ (F1)\end{tabular} & \begin{tabular}[c]{@{}c@{}c@{}}CoLA\\ (Matthews\\ Correlation)\end{tabular}& \begin{tabular}[c]{@{}c@{}c@{}}STS\\ (Pearson\\Correlation)\end{tabular} & \begin{tabular}[c]{@{}c@{}}Avg\\ (GLUE score)\end{tabular}\\ \hline
BERT-base & 768  & 3072 & 133.54M  & 82.1/83.0  & 86.6  & 88.8 & 64.6 & 90.1 & 85.6 & 46.5 & 87.9 & 79.8 (+4.5\%)\\ \hline
CVBERT-base & 768  & 1536 & 135.10M & 81.6/81.7 & 86.2 & 88.1 & 63.3 & 90.7  & 88.2 & 52.8 & 88.5 & 80.1 (+4.9\%)\\ \hline
QBERT-base & 768  & 1536 &  135.06M & 78.9/79.7 & 84.3  & 86.9 & 60.3 & 89.0 & 80.0 & 41.9 & 84.8 & 76.2 (+0.0\%)\\ \hline
QCLS-end2end & 768  & 1536 & 48.07M & 40.6/41.6 & 65.9 & 67.6 & 48.0 & 80.5 & 74.8 & 0.0 & -4.5 & 47.1 (-61.8\%)\\
QCLS-transformer-3L & 768  & 1536 & 71.30M & 55.0/55.1 & 65.8 & 72.6 & 45.1 & 80.6 & 65.8 & 3.3 & 2.4 & 49.8 (-53.0\%)\\
QCLS-transformer-6L & 768  & 1536 & 92.57M & 57.2/57.4 & 65.9 & 73.4 & 52.0 & 81.0 & 65.9 &  7.3& 11.3 & 50.8 (-50.0\%)\\
QCLS-transformer-12L & 768  & 1536 & 135.06M & 59.6/59.8 & 66.6 & 74.4 & 48.0 & 80.0 & 74.1 & 13.4 & 13.3 & 51.4 (-48.2\%)\\\hline

DisCoCat~\cite{lorenz2021qnlp} & & &  &  &  &  &  &50.9  &  & -2.0& \\ \hline\hline
\end{tabular}
\end{centering}
}
\end{table*}

\textbf{DisCoCat} is implemented with lambeq~\cite{kartsaklis2021lambeq}, an open-source, modular, extensible high-level Python library for experimental Quantum Natural Language Processing (QNLP). We apply \textbf{DisCoCat} to SST and CoLA, the text classification datasets in GLUE, and both training and prediction are simulated classically, with CAdamW and a batch size of 32. See App.~\ref{sec:discocat_sst2} for the source code. The remaining models are pre-trained on 8 Tesla V100 GPU cards, at a batch size of 512 and an initial learning rate of 1e-4. Fine-tuning models are executed on a single Tesla V100 card, at a batch size of 128 and an initial learning rate of 1e-3. All models are implemented in PyTorch 1.9.0.

\begin{figure}[ht!] 
\centering 
\includegraphics[width=0.47\textwidth]{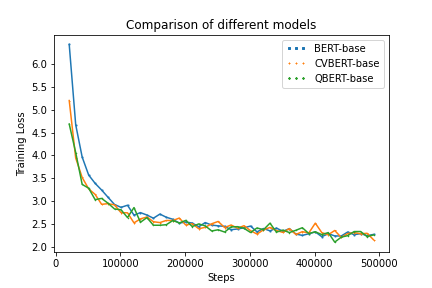}
\caption{Learning curves of real, complex and quantum BERT models.}
\label{fig:ablation_qmodels}

\end{figure}

\noindent \textbf{Overall Result}. Fig.~\ref{fig:ablation_qmodels} presents the learning curves of the three BERT models in pre-training, while Tab.~\ref{tab:qBERT} compares the accuracy performance of all above-mentioned models on GLUE dataset. \textbf{QBERT-base} has a close learning curve to \textbf{CVBERT-base} and \textbf{BERT-base} and a small 4.5\% and 4.9\% relative drop over the two models, indicating that the quantum-compatible settings on the network implementations have little harm to the performance.  More importantly, remarkable gaps in GLUE scores appear between \textbf{QBERT-base} and all end-to-end quantum classification models. By GLUE score, \textbf{QBERT-base} outperforms \textbf{QCLS-end2end} by 61.8\% and \textbf{QCLS-transformer} with a minimum gap of 48.2\%. Due to its low dimensionality, \textbf{DisCoCat} is even inferior to QCLS models by a large margin, and it is unfair to compare it with \textbf{QBERT-base}. However, the margin between \textbf{QBERT-base} and QCLS models does indicate the enormous benefit to the representation capacity of QNLP models brought about by the pre-trained quantum encoding. \\

\begin{figure}[!ht] 
\centering 
\includegraphics[width=0.45\textwidth]{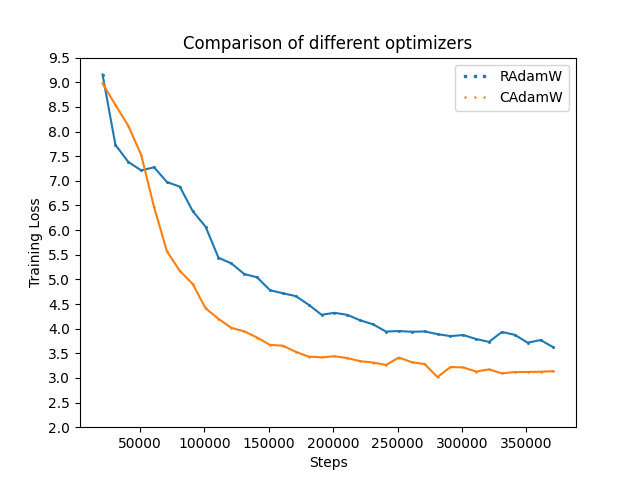}
\caption{Learning curves of real and complex Adam optimizers to train the same complex-valued model.}
\label{fig:ablation_optimizers}
\end{figure}

\noindent \textbf{CAdamW vs. RAdamW}. We compare the complex-valued AdamW optimizer with the original AdamW
optimizer (RAdamW) for real numbers. For a fair comparison, we use the two optimizers to pre-train the same complex-valued language model. As shown in Fig.~\ref{fig:ablation_optimizers}, CAdamW converges faster than RAdamW and to a lower loss in pre-training. Therefore, the improved AdamW optimizer is indeed superior to RAdamW for training complex-valued models in practice.

\noindent \textbf{Quantum Simulation}.
The experiment results are acquired by classical simulation. However, as demonstrated in App.~\ref{sec:app_qiskit}, the fine-tuning network can be converted to a quantum circuit can be implemented by the qiskit\footnote{https://qiskit.org/} toolbox, and both networks have identical behaviours. This implies that the reported values in the table can be obtained by hybrid classical-quantum training. The authentic classical-quantum hybrid simulation of QBERT is left for future work.
